# Anti-doping in Insulators and Semiconductors having Intermediate Bands with Trapped Carriers

Qihang Liu[1,2,*], Gustavo M. Dalpian[1,3] and Alex Zunger[1,*]

[1]*Renewable and Sustainable Energy Institute, University of Colorado, Boulder, Colorado 80309, USA*

[2]*Shenzhen Institute for Quantum Science and Technology and Department of Physics, Southern University of Science and Technology, Shenzhen 518055, China*

[3]*Centro de Ciências Naturais e Humanas, Universidade Federal do ABC, Santo André, SP, Brazil*

[*]*E-mail:* liuqh@sustc.edu.cn; Alex.Zunger@colorado.edu

**Abstract**

Ordinary doping by electrons (holes) generally means that the Fermi level shifts towards the conduction band (valence band) and that the conductivity of free carriers increases. Recently, however, some peculiar doping characteristics were sporadically recorded in different materials without noting the mechanism: electron doping was observed to cause a portion of the lowest unoccupied band to merge into the valance band, leading to a *decrease* in conductivity. This behavior we dub as "anti-doping" was seen in rare-earth nickel oxides $SmNiO_3$, cobalt oxides $SrCoO_{2.5}$, Li-ion battery materials and even MgO with metal vacancies. We describe the physical origin of anti-doping as well as its inverse problem – the "design principles" that would enable intelligent search of materials. We find that electron anti-doping is expected in materials having pre-existing trapped holes and is caused by annihilation of such "hole polarons" via electron doping. This may offer an unconventional way of controlling conductivity.

Doping of carriers into solids plays a crucial role both in controlling their physical properties (conductivity, superconductivity, metal-insulator transitions) via shifting of the Fermi level ($E_F$), and ultimately enables transport-based device technologies (electronics, spintronics, optoelectronics) [1,2]. Successful doping of insulators or semiconductors by electrons (holes) means that $E_F$ shifts towards the conduction band (valence band) and that the conductivity of free electrons (free holes) increases. The relationship $E_F(n)$ between the carrier density $n$ and the Fermi energy is textbook predictable [3] provided the density of states $D(\varepsilon)$ of the host solid (and hence its electronic structure) remains rigid (unperturbed by the doping process itself).

Recently, however, peculiar doping characteristics were noted in a number of disconnected cases, where electron doping was observed to significantly increase the band gap, and lead to a colossal *decrease* (several orders of magnitude) in conductivity. We will refer to such phenomenology as *"anti-doping"*. Such observations were recorded in materials systems such as rare-earth nickel oxides $SmNiO_3$ [4-6] and in ordered-vacancy cobalt oxides $SrCoO_{2.5}$ [7]. In contrast to normal doping that is governed by classic defect physics [1,8], anti-doping represents perhaps the most unprecedent extreme form of a non-rigid response of $D(\varepsilon)$ to doping, reversing entirely the expected trend – reducing, rather than increasing conductivity by doping. In sharp contrast to the well-established "unsuccessful doping" that is usually detrimental to applications, anti-doping paves a new route for band gap modulation and resistance switching, and thus promises new directions of doping-induced multiple functionalities such as fuel cells, electric field sensors, Li-ion battery materials, and optical devices [4-7,9]. Because of the disparity in properties of the systems where such peculiar doping characteristic was observed, it would be tempting to dismiss these observations as specific idiosyncrasy of specifically *complex or correlated systems*. In this work, we uncover a simple explanation of the hitherto peculiar electron anti-doping by the physics of "polaron annihilation", not due to electron localization by interelectronic repulsion. This understanding enables deliberate design and theoretical validation of new anti-doping compounds.

The behavior we call electron (hole) anti-doping illustrated in Fig. 1a and 1b is expected in insulators that prior to doping have, within their principal conduction-to-valence band gap, additional unoccupied (occupied) "intermediate bands" that have been split-off from the principal valence band (conduction band), containing "trapped" holes (electrons). Such intermediate bands are common in materials made of multi-valent elements. For example, *electron anti-doping* (see Fig. 1a) would occur in oxides where before doping the valence band made generally of doubly negative oxygen orbitals $O^{2-}$ splits off into the gap region an intermediate band [10], made of singly-negative (reduced) oxygen ions $O^{1-}$, containing polaronic trapped holes. We will illustrate four such materials in diverse oxides. Upon doping electrons, they do not populate the principal conduction band, but occupy instead the empty intermediate sub-bands, leading to compensation of the hole states in a cascade of steps. As a result, the split-off intermediate hole bands return to the valence home-base, merging with it, and thus increasing the occupied-to-unoccupied band gap and reducing conductivity. The complimentary process of *hole anti-doping* (see Fig. 1b) would occur in materials where before doping the conduction band splits off into the gap region an intermediate band containing trapped electrons, e.g., due to reduced metal ions. Upon doping holes, the split-off intermediate electron bands return to the conduction band home-base, merging with it, and thus increasing the occupied-to-unoccupied band gap as well. The concept of anti-doping compliments our textbook understanding of ordinary doping and could enable identification of systems and dopants that reduce conductivity upon demand. Furthermore, some anti-doping materials have the capacity to contain plenty of excessive electrons, indicating the potential to trigger dramatic multi-functional phase change by doping [5-7,11]. Particularly, some leading Li-ion battery reactions of lithiation involve precisely this effect: the insertion of Li *creates* anti-doping.

***Two types of hole states in insulators for electron anti-doping.*** –In the rest of the work we focus on electron anti-doping. In conventional metal-ligand compounds, the valence band is generally constructed from hybridized ligand orbitals in their nominal charge state (such as $O^{2-}$ in most ionic main group oxides MgO, CaO as well as in late transition metal oxides NiO, $Cu_2O$, ZnO [12]). Generally the unoccupied bands are composed of metal orbitals, but there is a small class of oxides in which empty oxygen states exist either as (a) symmetry-breaking (localized) hole polaron [13,14], or as (b) Bloch-periodic ligand-hole unoccupied bands [15-17]. The compounds with such conditions, as illustrated by Fig. 1a, could be considered as ideal host for anti-doping. The former cases are familiar from electron paramagnetic resonance of metal vacancies in main-group oxides [18,19], where holes created by metal vacancies (acting as acceptors) is localized on a *subset* of the oxygen ligands, pushing a defect-like, polaronic hole state from the valence band into the gap region. Periodic, ligand-hole bands, on the other hand, have the same symmetry as the cation orbitals and can be delocalized throughout the lattice rendering reduced ligand states (such as $O^{-1}$), as observed in some nickelates [16,17] and bismuthates [20], and another type of Li-ion cathode LiIrO$_3$ [21]. We next perform density functional theory (DFT) calculations to illustrate the anti-doping process in various materials with either pre-existing polaronic holes or ligand-hole band states. Note that currently available exchange-correlation functional ($E_{xc}$) in DFT usually fails systematically to predict localized polaron states where their formation is a fact because the self-interaction error often leads to an unrealistic delocalized wavefunction. A correction thus needs to be given to fulfill the so-called generalized Koopmans condition [13,14]

$$\Delta_{nk} = E(N-1) - E(N) + eig(N) = 0, \tag{1}$$

where $E(N-1) - E(N)$ denotes the total energy cost to remove an electron from the electron-doped system, and $eig(N)$ the single-particle energy of the highest occupied state in the electron-doped system. Therefore, for materials with pre-existing polarons we introduce a potential operator that acts only on the doping states to restore the generalized Koopmans condition (see Methods in Supplementary Materials [22]).

Before going to more peculiar compounds like nickelates, we first illustrate the concept of electron anti-doping in the simplest *s-p* orbital compound – MgO with a charge-neutral metal vacancy $V^0_{Mg}$. Early calculations of defects in oxides used semiclassical Mott-Littleton approach to polaron and more recent ones used more material-specific quantum-mechanical procedures [29]. As shown in Fig. 1c, for $V^0_{Mg}$ there are O hole polaron states localized inside the band gap, forming an empty intermediate band with two electrons' capacity. When adding an electron (forming $V^{-1}_{Mg}$), half of the intermediate band returns to the valence band, while doping two electrons (forming $V^{-2}_{Mg}$) causes the whole band (two symmetry-breaking polarons) to merge into the valence band (Fig. 1d and 1e). During the transition, the occupied-unoccupied band gap (marked as orange shadows in Fig. 1) increases, illustrating a reversible anti-doping via polaron formation and annihilation process with the doping capacity of two electrons (see also Fig. S1 [22]). Since the capacity of the intermediate band is quite limited by the low concentration of the defect centers (especially after annealing), the case of MgO is served as an introductory example of polaron annihilation. We will show below that a traditional polyanionic cathode LiFeSiO$_4$ also belongs to this anti-doping category in a wide doping range.

Anti-doping ligand-hole bands in extended bulk solids entails two conditions. The first condition is that the material have a *ligand-hole unoccupied band* (also referred to as "negative charge-transfer insulator" [15,30]) i.e., where the lowest unoccupied band has a significant O-$p$ component i.e., reduced oxygen $O^{1-}$. We will discuss in Figs. 3 and 4 examples of ligand-hole states later. The second condition is that these hole states form isolated intermediate bands inside the principal band gap, not connected via dispersion to the rest of the crystalline bands. In this case doped electrons that occupy such bands can exert a significant electrostatic effect without being delocalized onto the rest of the crystal (see Fig. 1a). Ordinary intermediate bands that are not ligand-hole states have been reported in a number of semiconductors used to enhance photovoltaic performance [10]. What we are looking for is the special case of ligand holes that are at the same time isolated intermediate bands. The two abovementioned conditions for anti-doping in extended, defect-free crystals set up "design principles" that act as filters to search the rare anti-doping bulk compounds. We next illustrate such cases.

*Electron anti-doping in Li-ion battery compounds.*–In cathode lithium-ion battery (LIB) compounds, lithiation and delithiation correspond to electron and hole doping, respectively. Transition-metal-ions in LIBs were initially regarded as the only source of redox activity providing charge-compensating electrons after lithiation or delithiation [31]. It was recently noted that the lowest unoccupied band in such compounds can be a trapped-hole state, with such ligand reduction raising the opportunity to boost the capacity and energy density of LIBs by combining both cationic (transition metal) and anionic (oxygen) redox processes within the same material [32,33]. We show that these processes represent anti-doping of isolated intermediate bands made of trapped holes by illustrating two cases of hole states affecting LIB: (a) where the intermediate band is localized, having a pure polaron-like character (Li$_x$FeSiO$_4$), and (b) the intermediate band is an extended ligand hole (Li$_x$IrO$_3$).

*Anti-doping with a polaronic intermediate band:* Li$_2$FeSiO$_4$ is a traditional polyanionic cathode material with a high capacity (corresponding to 1.86 Li per formula unit) that was attributed to the formal cation reaction $Fe^{4+} \rightarrow Fe^{2+}$ during the lithiation process from $FeSiO_4$ to $Li_2FeSiO_4$. However, previous DFT studies predicted a *metallic* state for $Li_xFeSiO_4$ during lithiation [34], in sharp contrast to the semiconducting phase reported by experiments [35]. Our calculations with self-interaction correction successfully confirm the semiconducting feature upon doping. More importantly, we suggested that anti-doping occurs in $Li_xFeSiO_4$ through the lithiation process with the annihilation of a localized, pure polaron-like hole, clarifying the previous contradiction and the nature of anion redox process. Fig. 2a-2c shows the projected density of states for three stages of lithiation in $Li_xFeSiO_4$, with $x = 0$, 0.5, and 1, respectively. For $x = 0$, instead of forcing Fe ion to have an unfavorable formal oxidation state of $Fe^{4+}$, two polaronic O states are formed inside the gap between valence band maximum and the Fe $e_g$ states around 3 eV (see Fig. 2d). Upon lithiation (electron doping), such empty O states are occupied gradually until no localized states exists inside the band gap. Therefore, through lithiation, only O states are active and the fundamental band gap increases from 1.0 eV, to 1.6 eV and 3.0 eV, for $x = 0$, 0.5 and 1 respectively, indicating an anti-doping behavior. We note that the antidoping in $Li_xFeSiO_4$ and in the metal vacancy of MgO (Fig. 1b-d) are very similar in that the host system has localized O polaronic states and the doping electron naturally occupied these states, rejoining the O reservoir in the valence band.

*Anti-doping with an extended ligand-hole intermediate band:* $Li_xIrO_3$ is illustrated in a ternary cathode material $LiIrO_3$ which exhibits anionic redox activity upon lithiation/delithiation [21]. The density of states in Fig. 3a shows that the lowest unoccupied bands form an intermediate band, composed by Ir-*d* and O-*p* states. The partial charge density of the intermediate band shown in Fig. 3c also confirms that $LiIrO_3$ has a ligand-hole band (negative charge-transfer insulator) with plenty of O ligand holes. Upon electron doping via lithiation, the number of electrons that can occupy this sub-band decreases; the band gap increases (see Fig. 3b), and the valence bands engulfs the O-p portion of these unoccupied sub-bands (while the contribution of Ir-d remains nearly the same), a process being the hallmark of the antidoping. This trend is further confirmed by the integrated charge density within a sphere centered at each Ir atom (see Fig. 3e), indicating that the actual charges residing around each Ir atom in $LiIrO_3$ and $Li_2IrO_3$ are very similar despite the formal valence states (5+ and 4+, respectively) being very different, a hallmark of the negative feedback "self-regulating response" [36], whereby the ligands rehybridize to donate electrons to the metal ion, protecting it from becoming overly positive.

*Electron anti-doping in perovskite-like nickelates.*–Both $SmNiO_3$ and $SrCoO_{2.5}$ were experimentally reported to show unconventionally increased band gap [4,7] and thus reduced conductivity upon electron doping, the phenomenology we attribute here to anti-doping behavior via electron compensation of their ligand hole bands. We next take $SmNiO_3$ as an example. Fig. 4a-4c show clear anti-doping behavior in $SmNiO_3$ with the bond-disproportionated low-T structure (the electron configuration of the two octahedra is $d^8$ and $d^8\underline{L}^2$, where $\underline{L}^2$ represents two ligand holes and both Ni site see an effective $Ni^{2+}$ charge). In the undoped compound there are two unoccupied intermediate bands close to the Fermi level: The one with lower energy is composed of Ni-3d and the surrounding O-2p, while the one with higher energy is almost purely composed by Ni-3d states. When the doping concentration reaches 1 e/Ni (Fig. 4c), the remaining portion of the lower intermediate band merges with higher-energy unoccupied bands, while the band gap increases from 0.5 eV to 2.1 eV. Fig. 4d and 4e show that there are significant O ligand states hybridizing with $Ni_1$ (small octahedral) for the undoped case, while the system turns to a positive charge-transfer insulator (no ligand holes) for 1 e/Ni doping, the experimentally doping limit. Fig. 4f and 4g show that upon doping the charge around Ni site remains nearly constant, indicating that the doping electron occupies selectively the O ligand-hole states in the lowest unoccupied band, dragging these states back to the valence band. Our explanation of the band gap increase in such as disproportionated system with constant $Ni^{2+}$ charges (no charge ordering) differs substantially from a previous explanation [6] that attributes the doping to the charge state transition of the cation $Ni^{3+}$ -> $Ni^{2+}$ based on an assumed insulating structure without bond disproportionation.

In $SmNiO_3$ electron-doping induced resistance modulation is experimentally observed. For example, the electron doping approaches were achieved by H insertion, Li insertion [4], as well as O vacancy [37]. Since excessive O vacancy would cause phase separation, the capacity to contain doping electrons is much smaller than that for H/Li insertion. Nevertheless, our anti-doping scenario explains the mechanism of such band-gap modulation generally, independent of the specific chemical identity of the dopant.

In summary, there are cases where the presence of doped free carriers can change the structure and symmetry of host solid lattice structure itself [38], or create self-trapped lattice polarons, or develop local, charge-compensating centers [39,40] that lead to the stagnation of $E_F$ ("pinning") even if carriers are added. Anti-doping represents perhaps the most extreme form of non-rigid response of $D(\varepsilon)$ to doping, reversing entirely the expected trend—reducing, rather than increasing conductivity by doping. We point out here that just like Li-ion compounds, or MgO with metal vacancy, electron anti-doping is rather general effect not related to correlation-induced phenomenology but rather the fact that valence band derived O ligand-hole states are easily electron doped (no electron repulsion), causing the empty O states to return to valence bands. In parallel, we expect hole anti-doping in materials where prior to doping there is an intermediate band split-off from the main conduction band containing trapped holes. Upon hole doping the occupied intermediate band will merge into the principle conduction band thus increase the occupied-to-unoccupied gap. This could be the case for compounds with reduced cations (e.g., $Ti^{+Q}$ with a Ti charge $Q$ below 4+ due to oxygen vacancies), as in the Magnelli phases of $TiO_2$ [41].

**Acknowledgements**

The work of A.Z., Q.L. and G.M.D. at the University of Colorado Boulder was supported by the U.S. Department of Energy, Office of Science, Basic Energy Sciences, Materials Sciences and Engineering Division under Grant No. DE-SC0010467. Q.L. was also supported by Guangdong Innovative and Entrepreneurial Research Team Program (Grant No. 2017ZT07C062). G.M.D. also thanks financial support from Brazilian agencies FAPESP and CNPq. Q.L. thanks Dr. Jiaxin Zheng, Dr. Guangfu Luo and Dr. You Zhou for helpful discussions. This work used resources of the National Energy Research Scientific Computing Center, which is supported by the Office of Science of the U.S. Department of Energy under Contract No. DE-AC02-05CH11231.

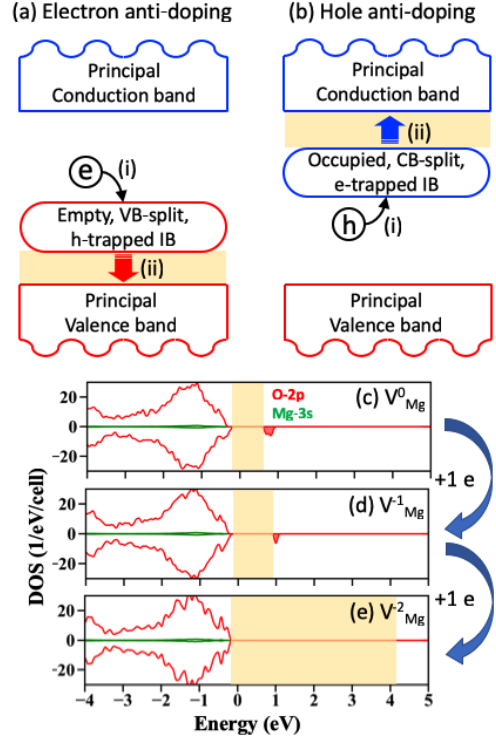

Fig. 1: Schematic illustration of (a) electron anti-doping and (b) hole anti-doping. Prior to doping in (a) the compound has an empty, valence-band-split (VB-split) intermediate band (IB) with trapped holes, whereas in (b) it has an occupied, conduction-band-split (CB-split) IB with trapped electrons. Doping (a) by electrons [step (i)] causes electron-hole recombination that results in shifting of the IB towards the principal VB [step (ii)]. This increases the band gap of the undoped system (orange shadow) and reduces conductivity. Analogous process occurs in (b) for hole anti-doping. (c-e) Illustration of electron anti-doping in DFT calculation on Mg vacancy in MgO, showing the density of states (DOS) of (c) $V^0_{Mg}$, (d) $V^{-1}_{Mg}$ and (e) $V^{-2}_{Mg}$.

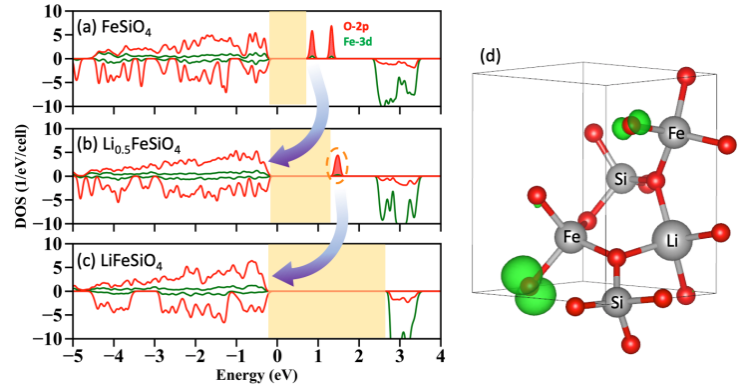

Fig. 2: (a-c) Density of states (DOS) of (a) $FeSiO_4$ and (b) $Li_{0.5}FeSiO_4$ and (c) $LiFeSiO_4$. The arrows indicate the intermediate band being swallowed by the valance band upon electron doping. (d) Module squared wavefunction (green isosurface) of the unoccupied intermediate bands [circled in panel (b)] for $Li_{0.5}FeSiO_4$. O atoms are marked by red.

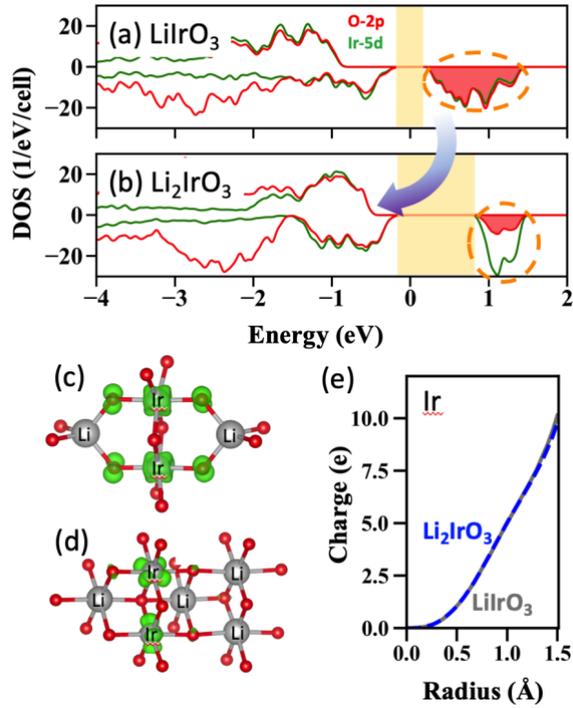

Fig. 3: (a,b) Density of states (DOS) of (a) LiIrO$_3$ and (b) Li$_2$IrO$_3$. The arrow indicates a portion of the intermediate band being swallowed by the valance band upon electron doping. (c, d) Module squared wavefunction (green isosurface) of the unoccupied intermediate bands [circled in panel (a) and (b), respectively] for (c) LiIrO$_3$ and (d) Li$_2$IrO$_3$. O atoms are marked by red. (e) Total charge density integrated in a sphere centered in an Ir atom for LiIrO$_3$ (grey) and Li$_2$IrO$_3$ (blue) as a function of radius, showing almost constant charge density on the nominal Ir$^{5+}$ and Ir$^{4+}$ cations.

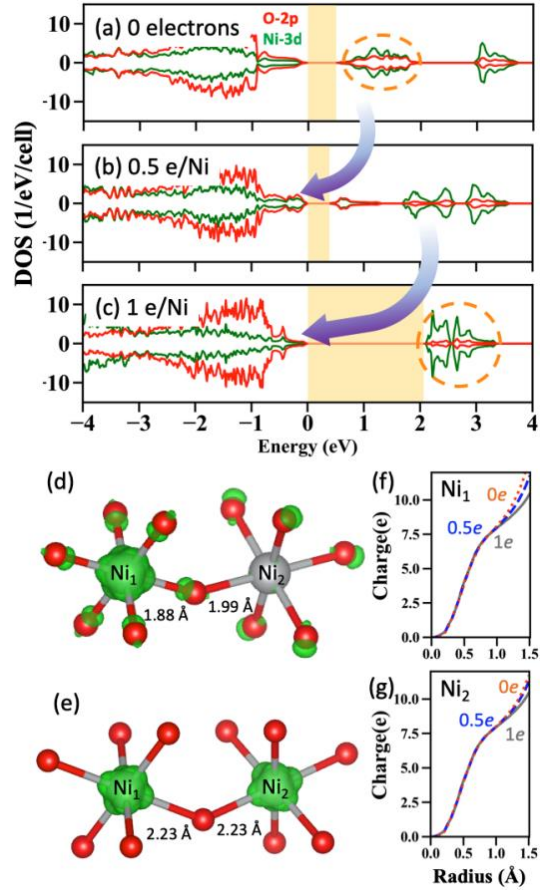

Fig. 4: (a-c) Density of states (DOS) of SmNiO$_3$ for (a) undoped (b) doped by 0.5 e/Ni and (c) doped by 1 e/Ni atom. The arrows indicate a portion of the intermediate band being swallowed by the valance band upon electron doping. (d, e) Module squared wavefunction (green isosurface) of the unoccupied intermediate bands [circled in panel (a) and (c), respectively] for (d) undoped and (e) doped by 1 e/Ni. O atoms are marked by red. (f, g) Total charge density integrated in a sphere centered in the Ni atom inside (f) the small octahedra (Ni$_1$) and (g) the large octahedra (Ni$_2$) as a function of radius.